\documentstyle[12pt,prl,aps,psfig]{revtex}

\begin{document}
\draft
\title{Spin Relaxation Caused by Thermal Excitations of High Frequency Modes of Cantilever Vibrations}
\author{G.P. Berman$^1$, V.N. Gorshkov$^{1,2}$, D. Rugar$^{3}$, and V.I. Tsifrinovich$^4$}
\address{$^1$ Theoretical Division, Los Alamos National
Laboratory, 
Los Alamos, New Mexico 87545}
\address{$^2$ Department of Physics, Clarkson University, Potsdam, NY 13699}
\address{$^3$ IBM Research Division, Almaden Research Center, 650 Harry Rd., San Jose, California 95120-6099}
\address{$^4$ IDS Department, Polytechnic University, Brooklyn, NY 11201}
\maketitle
\vspace{5mm}
\begin{abstract}
We consider the process of spin relaxation in the oscillating cantilever-driven adiabatic reversals technique in magnetic resonance force microscopy. We simulated the  spin relaxation caused by thermal excitations of the 
high frequency cantilever modes in the region of the Rabi frequency of the spin sub-system. The minimum relaxation time obtained in our simulations is greater but of the same order of magnitude as one measured in recent experiments. We demonstrated that using a cantilever with nonuniform cross-sectional area may significantly increase spin relaxation time.
\end{abstract}
\section{Introduction}
The method of oscillating cantilever-driven adiabatic reversals (OSCAR) is a promising method for spin detection in magnetic resonance force microscopy (MRFM) \cite{1,2}. Just like the ordinary MRFM technique \cite{3,4}, the magnetic moment of a sample driven by a resonant microwave magnetic field interacts with a ferromagnetic particle. If the ferromagnetic particle is attached to the cantilever tip, the magnetic moment of the sample changes the cantilever vibrations. This change can be detected through measurement of the cantilever vibration parameters. In the OSCAR technique the cantilever is driven by an external force with a feedback loop designed to keep the amplitude constant. The cantilever vibrations together with the resonant microwave magnetic field cause a cyclic adiabatic inversion of the magnetic moment of the sample. In turn, this cyclic inversion causes a shift of the cantilever vibration frequency, which is measured.

The main challenge in the OSCAR technique is the sharp decrease of the MRFM relaxation time when the distance between the ferromagnetic particle and the sample is below 1 micrometer \cite{1}. Thus, an important problem of the MRFM technique is to understand the nature of the spin relaxation in the submicron region and to find a way to reduce the relaxation rate. One of the possible mechanisms involved in MRFM relaxation in the submicron region is thermal currents in the metallic ferromagnetic particle \cite{1}. Another possible mechanism is thermal excitation of the high frequency modes of the cantilever vibrations \cite{5}. 

In this paper, we explore this second mechanism. We show that the thermal excitations of the high frequency modes may provide a noticeable contribution to the experimentally observed spin relaxation of the MRFM signal. We also show that the use of a nonuniform cantilever can reduce the spin relaxation rate caused by this mechanism.

\begin{figure}[t]
\centerline{\psfig{file=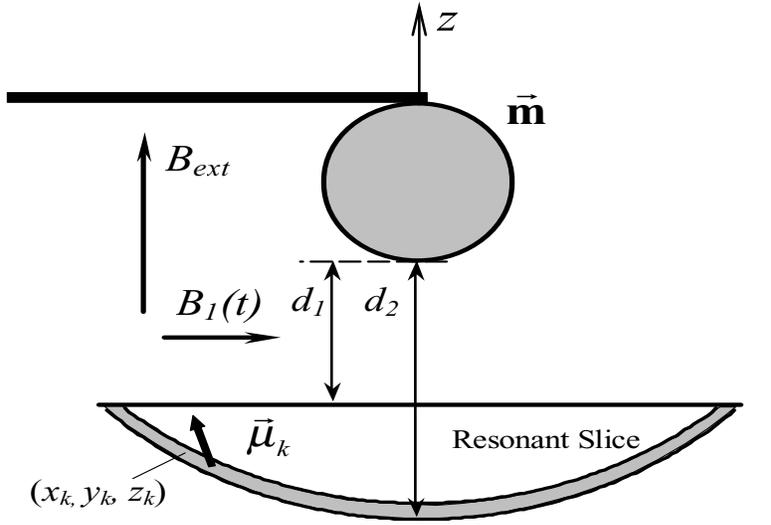,width=11cm,height=9cm,clip=}}
\vspace{4mm}
\caption{The geometry of the OSCAR technique.}
\label{fig:1}
\end{figure}

\section{Model and equations of motion}

The geometry of the problem is shown in Fig. 1.
 This geometry is essentially the same as that considered in \cite{2}. However, now we consider not just a single magnetic moment below the cantilever tip, but a uniform distribution of the magnetic moments across the sample which corresponds to the experimental conditions in \cite{1}. The ferromagnetic particle attached to the cantilever tip effectively interacts with all magnetic moments inside the resonant slice where the Larmor frequency matches the frequency of the microwave magnetic field, $B_1$. (The effective thickness of the resonant slice equals twice the amplitude of the cantilever vibrations.)

The tip vibrations associated with the fundamental mode of the cantilever can be described as a motion of an oscillator with a quality factor $Q$:
$$
\ddot z_c+z_c+\dot z_c/Q=f(\tau),\eqno(1)
$$
$$
f(\tau)=\sum_k\eta_k\mu_{kz},
$$
$$
\eta_k={{\mu_0}\over{4\pi}}{{3m\mu}\over{k_cz^5_0}}{{\tilde z_k(5{\tilde z}_k^2-3{\tilde r}^2_k)}\over{{\tilde r}^7_k}}.
$$
Here we used the dimensionless quantities: $\tau=\omega_ct$, where $\omega_c$ is the ``unperturbed'' cantilever frequency of the fundamental mode; $z_c$ is the $z$-coordinate of the center of ferromagnetic particle in the units of the cantilever amplitude $z_0$; $x_k$, $y_k$, $z_k$ are the coordinates of the $k$-th magnetic moment of the sample in the same units; $\tilde z_k=z_k-z_c$; ${\tilde r}^2_k=x_k^2+y_k^2+{\tilde z}_k^2$. $\mu_0$ is the permeability of free space; ${\vec \mu}_k$ is the $k$-th magnetic moment in units of its magnitude $\mu$ (which is the same for all magnetic moments); $k_c$ is the cantilever spring constant, $m$ is the magnetic moment of the ferromagnetic particle. The function $f(\tau)$ describes the action of the magnetic moments in the resonant slice on the cantilever.

The motion of the $k$-th macroscopic magnetic moment in the rotating reference frame  is given by
$$
\dot \mu_{kx}=-\Delta_k \mu_{ky},\eqno(2)
$$
$$
\dot \mu_{ky}=\Delta_k \mu_{kx}-\varepsilon\mu_{kz},
$$
$$
\dot \mu_{kz}=\varepsilon\mu_{ky}.
$$
In Eqs. (2) the following notation was used:
$$
\Delta_k=(\gamma B_{ext}-\omega)/\omega_c+{{\mu_0}\over{4\pi}}{{\gamma m}\over{\omega_cz^3_0}}{{3\tilde z^2_k-{\tilde r}^2_k}\over{{\tilde r}^5_k}},\eqno(3)
$$
$$
\varepsilon=\gamma B_1/\omega_c.
$$
Here $\gamma$ is the electron gyromagnetic ratio in the sample; $B_{ext}$ is the external static magnetic field; $B_1$ and $\omega$ are the amplitude and frequency of the rotating {\it rf} magnetic field. Note, that $\Delta_k$ is the $z$-component of the rotating-frame effective field ${\vec B}^k_{eff}$ (in units $\omega_c/\gamma$), and $\varepsilon$ is the $x$-component of the effective field in the same units, ${\vec B}^k_{eff}=(\varepsilon,0,\Delta_k)$.

The upper and the lower boundaries of the resonant slice are determined by the equation
$$
\Delta_k=0,~at~z_c=\pm 1.
$$
In our numerical experiments, the magnetic moments have been distributed uniformly inside the resonant slice. We used the following initial conditions
$$
z_c=-1,~\mu_{kz}=1,\eqno(4)
$$
i.e. the magnetic moments are oriented approximately along the effective field in the rotating frame. To model the feedback technique in the OSCAR MRFM, our computer algorithm increased the value of $z_c$ to 1 
every time the cantilever passed the upper point. The period of the cantilever oscillations was determined as the time interval between the instants of the $z_c$ maximum values. 

To model the thermal excitations of the high frequency cantilever modes we use the following approach. The cantilever energy $E$ can be written as a sum of the energies of all vibrational modes. The high frequency modes can be described approximately by the formula \cite{6}
$$
\omega_n={{c_nt_c}\over{l^2}}\Bigg({{Y}\over{12\rho}}\Bigg)^{1/2},\eqno(5)
$$
$$
c_n=[\pi(n-0.5)]^2,~n>1,
$$
where $t_c$ is the thickness of the cantilever, $l$ is its length, $Y$ is  Young's modulus, and $\rho$ is the density of the cantilever. Combining the expressions
$$
{{m_c\omega^2_na^2_n}\over{2}}=k_BT,~\omega^2_c=4k_c/m_c,\eqno(6)
$$
where $m_c$ and $k_c$ are the mass and the spring constant of the cantilever, we estimate the thermal amplitude of the $n$-th mode to be
$$
a_n={{1}\over{\Omega_n}}\Bigg({{k_BT}\over{2k_c}}\Bigg)^{1/2},~\Omega_n=\omega_n/\omega_c.\eqno(7)
$$
The amplitude of the cantilever tip oscillations, $b_n$, for any high frequency mode, $n$, is twice the amplitude of the mode \cite{6}: 
$$
b_n=2a_n= (2k_BT/k_c\Omega^2_n)^{1/2}.
$$
To describe the influence of the noise on the spin dynamics we replace the coordinate $z_c$ with $z_c+\delta z_c$ in the expression for $\tilde z_k$ and $\tilde r_k$ in (3), where
$$
\delta z_c=\sum_n{{b_n}\over{z_0}}\cos(\Omega_n\tau+\Psi_n),\eqno(8)
$$
and $\Psi_n$ being a random phase.

\section{Results of numerical simulations}

We solved the system of equations (1), (2) changing $z_c\rightarrow z_c+\delta z_c$ in Eqs. (2) according to (8). (We did not take into account the influence of the thermal noise on $\eta_k$ in Eqs. (1). As our simulations demonstrate, the influence of the thermal noise does not cause a significant direct contribution to the cantilever vibrations through the parameter $\eta_k$, but causes a dephasing of magnetic moments in the resonant slice.) The number of magnetic moments in the resonant slice was 200 ($1\le k\le 200$). Our simulations show that the results do not change significantly when this number is increased to 400.

The parameters  in our numerical simulations were taken from 
experiment \cite{1}:
$$
B_{ext}=140~{\rm mT},~k_c=0.014~{\rm N}/{\rm m},~\omega_c/2\pi=21.4~{\rm kHz},~z_0=28~{\rm nm},\eqno(9)
$$
$$
Q=2\times 10^4,~{\rm m}=1.5\times 10^{-12}~{\rm J/T}.
$$
The distance between the bottom of the ferromagnetic particle and the surface of the sample, $d_1$, (see Fig. 1) was $d_1=700$ nm, and the radius of the ferromagnetic particle was equal to $d_1$.
The average magnetization of the sample was 0.89 A/m. The rotating {\it rf} field amplitude, $B_1$, in experiments was no less than 0.15 mT. The cantilever temperature, T, did not exceed 20 K. (For these values of parameters the value of $d_2$ in Fig. 1 was found to be $d_2=875$ nm, and the value of the magnetic field gradient at the center of the resonant slice was $1.4\times 10^5$ T/m.)

We have studied the decay of the OSCAR signal, $\Delta T/\Delta T_0$, where $\Delta T$ is the shift of the period of the cantilever vibrations due to the influence of the magnetic moments of the sample, and $\Delta T_0$ is the initial value of $\Delta T$. In our model, the decay of the OSCAR signal is associated with cantilever noise near the Rabi frequency $\omega_R=\gamma B_1$ which is the smallest resonant frequency in the rotating reference frame.

High frequency cantilever modes may generate significant noise near the Rabi frequency causing a deviation of the magnetic moments from the direction of the effective field and a nonuniform dephasing between magnetic moments.

We write the amplitude of the cantilever tip thermal vibrations associated with the $n$-th mode as $b_n=b_R\varepsilon/\Omega_n$, where
$$
b_R=\varepsilon^{-1}(2k_BT/k_c)^{1/2}.
$$

The phases $\Psi_n$ in (8) were changed randomly between $0$ and $2\pi$ with random time intervals, $\tau_\Psi$, between two successive changes of the phase. In particular, we put $\tau_\Psi=N\tau_0$, where $\tau_0$ takes random values between $2\pi/1.2\varepsilon$ and 
$2\pi/0.8\varepsilon$, and $N$ is a free parameter of the model.

Our simulations show that the decay of the OSCAR signal is almost 
independent of $N$ for $N<1000$. (See Fig. 2.) We have found that for $N<1000$ the signal decay can be approximately described by an exponential function with relaxation time, $\tau_m$ (Fig. 2). 

\begin{figure}[t]
\centerline{\psfig{file=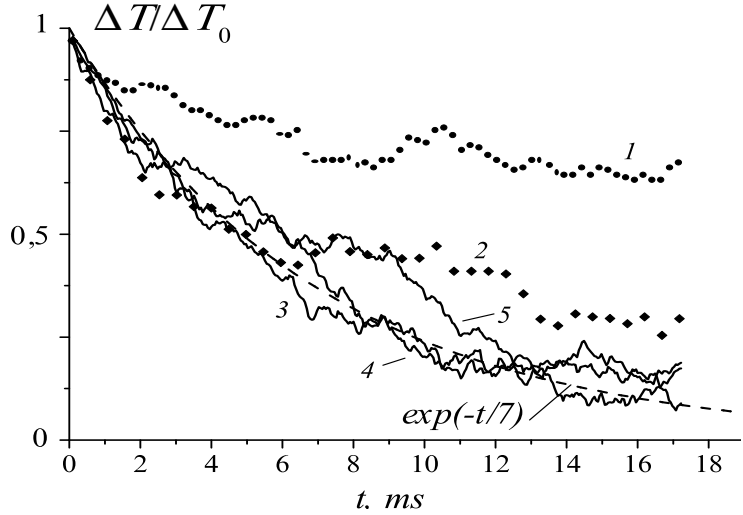,width=11cm,height=8cm,clip=}}
\vspace{4mm}
\caption{Decay of the OSCAR signal for different values of the parameter $N$. Curves 1-5 correspond to $N=10^5, 10^3, 100, 10$ and $2$. The amplitudes: $b_R=5$pm, $z_0=15$nm, and $\varepsilon=390$.}
\label{fig:2}
\end{figure}

The relaxation time quickly decreases with an increase of the characteristic noise amplitude at the Rabi frequency, $b_R$ which, in turn, is proportional to $\sqrt{T}/\varepsilon$. Taking the minimum value of $\varepsilon$ ($\varepsilon=195$, i.e. $B_1=0.15$mT) and the maximum value of $T$ ($T=20$K) from the range of parameters \cite{1}, we obtain the maximum value of $b_R$, $b_R=1$pm. The corresponding minimum value of the relaxation time is approximately $\tau_m\approx 570$ms. The experimental value found in \cite{1}, $\tau_m=68$ms, is smaller but of the same order of magnitude as our minimum theoretical value. Thus, we conclude that the thermal vibrations of the high frequency modes provide a noticeable contribution to the spin relaxation.

To study the spin relaxation we also considered values of parameters which provide relatively small relaxation time. This allowed us to reduce the simulation time and to determine the characteristic scaling properties of the relaxation process. As an example, Fig. 3 shows the decay of the OSCAR signal for various values of the cantilever amplitudes. One can see that the relaxation time $\tau_m$ decreases when the cantilever amplitude decreases. (See curves $a$ and $c$ in Fig. 3.) Note that if we take into consideration spins near the resonant slice, the value of $\tau_m$ slightly increases. (Compare curves $b$ and $c$ in Fig. 3.)

\begin{figure}[t]
\centerline{\psfig{file=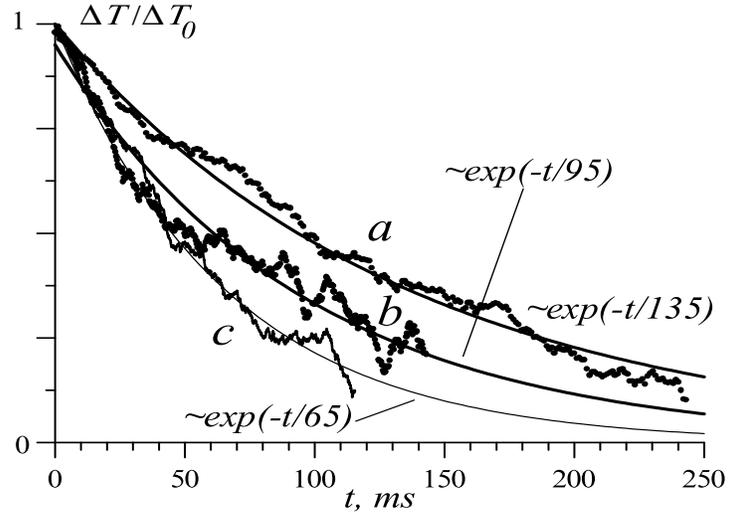,width=11cm,height=8cm,clip=}}
\vspace{4mm}
\caption{Decay of the OSCAR signal for various values of the cantilever amplitude $z_0$;  $b_R=1$pm, and $\varepsilon=390$. Curve $a$ corresponds to $z_0=15$nm; curves $b$ and $c$ correspond to $z_0=7.5$nm. For curve $b$ we took into consideration spins close to the resonant slice.}
\label{fig:3}
\end{figure}
\begin{figure}[t]
\centerline{\psfig{file=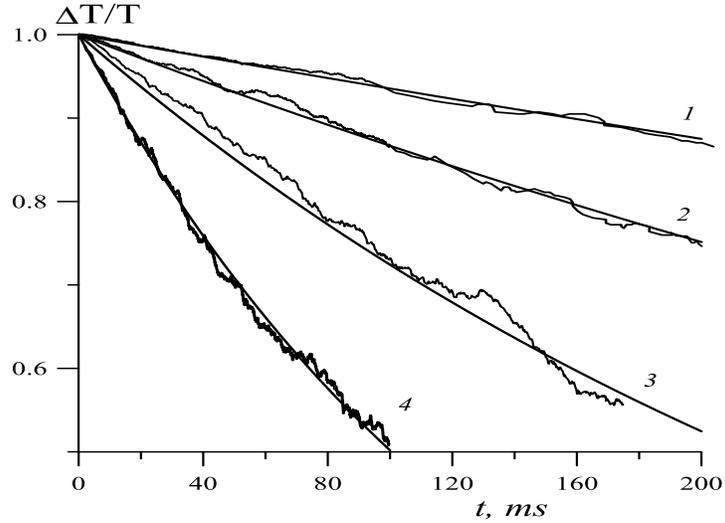,width=11cm,height=8cm,clip=}}
\vspace{4mm}
\caption{Decay of the OSCAR signal for $z_0=28$nm and various values of the temperature $T$ and rotating magnetic field $B_1$. Curves 1-3 correspond to $\varepsilon=390$ ($B_1=0.3$mT) and $T=20$K, $40$K and $80K$. Curve 4 corresponds to $\varepsilon=195$ ($B_1=0.15$mT) and $T=80$K. The relaxation time in curves 1-4 is 1500ms, 700ms, 310ms, and 145ms.}
\label{fig:4}
\end{figure}

Fig. 4 shows the decay of the OSCAR signal for various values of the temperature and rotating magnetic field. One can observe the expected decrease of the relaxation time $\tau_m$ with an increase of the temperature or decrease of the rotating magnetic field.

Fig. 5 shows the decay of the OSCAR signal for various numbers of 
high frequency  modes taken into consideration. One can see that the relaxation time $\tau_m$ slightly decreases when the number of high frequency cantilever modes increases from 2 to 22. A further increase of the number of modes does not essentially change  the decay rate. 

Our simulations show that the OSCAR signal generally consists of two parts: (a) a regular part which we studied in our paper and (b) a random part which could be observed after the decay of the regular signal. Numerical analysis of the random part of the OSCAR signal requires large computational times, and will be studied elsewhere.

\begin{figure}[t]
\centerline{\psfig{file=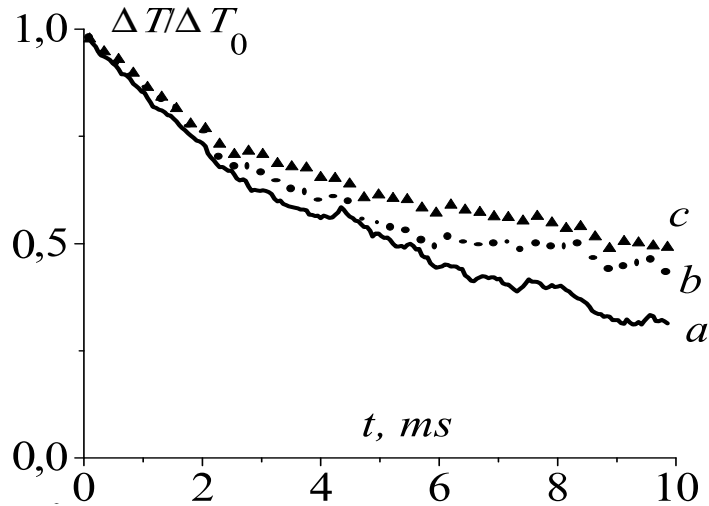,width=11cm,height=8cm,clip=}}
\vspace{4mm}
\caption{Decay of the OSCAR signal for $b_R=5$pm, $z_0=15$nm, and various numbers of the high frequency modes taken into consideration. The lowest of high frequency cantilever modes is the mode with the frequency closest to the Rabi 
frequency $8.4$MHz ($\varepsilon=390$). Curves $a$, $b$, and $c$ correspond to 22, 3, and 2 high frequency cantilever modes including the lowest one.}
\label{fig:5}
\end{figure}

\section{Reduction of the relaxation rate}

In this section we discuss how to reduce the spin relaxation rate caused by thermal excitations of the high frequency cantilever modes. For this purpose we consider a cantilever with nonuniform thickness.

The mechanical energy of the cantilever can be written as
$$
E={{1}\over{2}}\int^l_0\Bigg[S\rho\Bigg({{\partial z}\over{\partial t}}\Bigg)^2+YI\Bigg({{\partial^2z}\over{\partial x^2}}\Bigg)^2\Bigg]dx.\eqno(10)
$$
Here $S$ is the cross-sectional area, $I=wt^3_c/12$, and $w$ is the width of the cantilever. The equation for the cantilever motion without an external force and damping is given by
$$
S(x)\rho{{\partial^2z}\over{\partial t^2}}=-Y{{\partial^2}\over{\partial x^2}}\Bigg(I(x){{\partial^2z}\over{\partial x^2}}\Bigg).\eqno(11)
$$
Using a new variable, $\zeta=z\sqrt{S(x)}$, we obtain the equation of motion in the form
$$
\rho{{\partial^2\zeta}\over{\partial t^2}}=-{{Y}\over{\sqrt{S(x)}}}{{\partial^2}\over{\partial x^2}}\Bigg(I(x){{\partial^2}\over{\partial x^2}}{{\zeta}\over{\sqrt{S(x)}}}\Bigg).\eqno(12)
$$
The general solution of the equation (12) can be written as
$$
\zeta(x,t)=\sum_na_nf_n(x)\exp(-i\omega_nt).\eqno(13)
$$
The eigenfunctions $f_n(x)$ are normalized in the following way:
$$
\int^l_0S(x)f_n(x)f_m(x)dx=V\delta_{nm},\eqno(14)
$$
where $V=\int^l_0S(x)dx$ is the cantilever volume. The equation for the eigenfunctions is 
$$
\rho\omega^2_nf_n(x)=-{{Y}\over{\sqrt{S(x)}}}{{\partial^2}\over{\partial x^2}}\Bigg(I(x)
{{\partial^2}\over{\partial x^2}}{{f_n(x)}\over{\sqrt{S(x)}}}\Bigg).\eqno(15)
$$
We express the solution of Eq. (15) in the form
$$
f_n(x)=\sum_{k=1}^{k_{max}}c_{kn}\varphi_k(x),\eqno(16)
$$
where $\varphi_k(x)$ is the $k$-th eigenfunction for a cantilever with uniform cross-sectional area \cite{6}, satisfying the equation:
$$
S\rho\omega^2_n\varphi_n=YI{{\partial^4\varphi_n}\over{\partial x^4}},\eqno(17)
$$
$$
\int^l_0\varphi_n(x)\varphi_m(x)dx=l\delta_{nm}.
$$
Taking $k_{max}=70$, we use a standard procedure to find the coefficients $c_{kn}$ and frequencies $\omega_n$. 
Finally, using the same approach as for the uniform cross-sectional area we can find thermal vibrations of the cantilever tip, $\delta z_c$, in (8) with new values of $\Omega_n$ and $b_n$.

We found that it is possible to reduce the relaxation rate using a nonuniform thickness of the cantilever. (See also \cite{5}.) We considered a cantilever whose thickness increases near the location of the ferromagnetic particle:
$$
\delta t_c(x)=\delta t_c(l)\exp[-(x-l)^2/a^2].\eqno(18)
$$
For the value $a/l\sim 0.01$, we obtained an almost tenfold increase in the relaxation time.

We also considered the case when the ferromagnetic particle was located away from the tip of the cantilever. We found that in this case the increase of the cantilever thickness near the location of the ferromagnetic particle causes a significant increase of the relaxation time.
The details of these results will be published elsewhere.

\section{Conclusion}

We simulated the spin relaxation in OSCAR MRFM technique caused by the thermal excitation of the high frequency cantilever modes. Using a range of experimental parameters \cite{1} we obtained the minimal relaxation time 570 ms which is greater than but of the same order of magnitude as the experimental value 68 ms. We demonstrated that an increase of the cantilever thickness at the location of the ferromagnetic particle can significantly increase the relaxation time.

Finally, we would like to mention two assumptions of our computational model: We assumed 1) that the mass of the ferromagnetic particle was small compared to the cantilever mass, and 2) that the ferromagnetic particle has a spherical shape. We are now simulating a modified cantilever-spin model which is not restricted by these two assumptions. 

\section*{Acknowledgments}

We thank G.D. Doolen and  D. Mozyrsky for discussions. This work  was supported by the Department of Energy under the contract W-7405-ENG-36 and DOE Office of Basic Energy Sciences, and by the DARPA Program MOSAIC.

{}
\end{document}